\documentstyle[twocolumn,prl,aps,epsfig,floats]{revtex}
\tighten

\def\beq {\begin{equation}}
\def\eeq {\end{equation}}
\def\beqa {\begin{eqnarray}}
\def\eeqa {\end{eqnarray}}

\begin{document}
\title{Domain Wall Production During Inflationary Reheating}
\author{M. F. Parry$^1\,$\cite{mail1} and A. T.
Sornborger$^2\,$\cite{mail2}}
\address{~\\
$^1$Physics Department, Brown University, Providence, RI 02912, USA \\
$^2$NASA/Fermilab Astrophysics Group, Fermilab National Accelerator 
Laboratory, Box 500, \\Batavia, IL 60510-0500, USA}
\maketitle

\begin{abstract}
\noindent
We numerically investigate the decay, via parametric resonance, of the
inflaton with an $m^2 \phi^2$ potential into a scalar matter field with a
symmetry breaking potential. We consider the case where symmetry breaking
takes place during inflation. We show that when expansion is not taken
into account symmetry restoration and non-thermal defect production during
reheating is possible. However in an expanding universe the fields do not
spend sufficient time in the instability bands; thus symmetry restoration
and subsequent domain wall production do not occur. 
\end{abstract}

\smallskip

 
\noindent{BROWN-HET-1121, April 1998 \\
FERMILAB-Pub-98/135-A}
\narrowtext

\vskip0.8cm \noindent {\bf Introduction}

The realisation that the decay of the inflaton might occur explosively,
during a stage dubbed ``pre-heating'' \cite{tb90,kls94,stb95} and lead to a
universe far from thermal equilibrium, has a number of important
ramifications. The most obvious is that the temperature of the universe,
after thermalisation, may be much higher than under the old model of
reheating. Another is the recent construction of a model for baryogenesis
\cite{krt98} which takes advantage of the out-of-equilibrium nature of the
universe shortly after pre-heating. We, on the other hand, will be
interested in the possibility of non-thermal phase transitions after
pre-heating, and the possible production of topological defects.
 
This suggestion was first made by Kofman et al. \cite{kls96}, and Tkachev
\cite{t96}. The essential idea is that pre-heating, via the mechanism of
parametric resonance, typically gives rise to very large fluctuations in
the fields and it is these that may restore a symmetry which has been
broken by the time inflation ends. The important point is that symmetry
restoration may occur even if the symmetry breaking scale is higher than
the final reheat temperature. Then as the universe subsequently expands
and cools, phase transitions will result in the creation of topological
defects. This would be very intriguing. For a start we might find the
re-emergence of a problem inflation was designed to solve, namely the
monopole problem. The creation of domain walls would also be potentially
problematic. However the production of strings may be a desirable feature
of such scenarios, in that strings can be a seed for structure formation.

Khlebnikov et al. \cite{kklt98} have recently demonstrated that
non-thermal defect production indeed occurs during inflationary reheating. 
They considered a model in which the inflaton field $\phi$ has a double
well potential and is coupled to massless scalar matter fields $X_i,
i=1\ldots N$. Similar results (obtained from 2-D simulations) have also
been shown by Kasuya and Kawasaki \cite{kk98}. 

We study a different model. In our work the inflaton $\phi$ has a unique
minimum and decays via parametric resonance into a scalar matter field
$\chi$ with the discrete symmetry $\chi \rightarrow -\chi$. That is, we
are interested in the question as to whether domain walls are produced.
(Note that we are not considering a two-field model of inflation.) We
assume that symmetry breaking occurs during inflation, for if it occurs
afterwards the symmetry will then be restored a number of times by the
oscillation of the inflaton alone, without any need for parametric
resonance, and defect production will almost certainly ensue. This was
pointed out by Kofman et al. in their original paper \cite{kls96}, where
this model was first examined in the context of parametric resonance and
defect production, and by Kofman and Linde in \cite{kl87}. 

We investigate both the expanding and non-expanding cases and we show that
symmetry restoration and defect production only occur in the non-expanding
case. However the defects are not stable. Our work is based on 3-D lattice
simulations, though we also understand the results from an analytical
point of view.

The paper is set out as follows. We begin with a detailed explanation of
the model including necessary constraints on the model. The second section
is devoted to the analytical understanding of the model and predictions
for the full simulation. After this we highlight the workings of the
numerical code and present the results from our simulations. The final
section is a brief summary of our study and conclusion. 

\vskip 0.5cm \noindent{\bf 1. Model}

We consider the simplest model of inflation and reheating: the inflaton
$\phi$ has potential $\frac{1}{2}m^2\phi^2$ and it decays into a second
scalar field $\chi$. We let $\chi$ have the simplest potential which
allows symmetry breaking: $\frac{1}{4}\lambda(\chi^2-\chi_0^2)^2$,
and suppose it interacts with $\phi$ through the interaction term: 
$\frac{1}{2}g^2\phi^2\chi^2$. Thus the full potential for the two-field
theory is: 

\beq \label{pot}
V(\phi,\chi) = \frac{1}{2}m^2\phi^2 +
\frac{1}{4}\lambda(\chi^2-\chi_0^2)^2 + \frac{1}{2}g^2\phi^2\chi^2 \
\eeq

The most important constraint on our model is the requirement that the
symmetry breaking occurs during inflation. This is equivalent to demanding
the interaction term in (\ref{pot}) not lift the degeneracy of the
potential in the $\chi$ direction. As $\phi$ is homogeneous to a very
good approximation for initial times, we may write $\phi \approx \phi(t)$
so that the effective potential for $\chi$ is:

\beq \label{poteff}
V^{eff}_{\chi} = \frac{1}{4}\lambda(\chi^2-\chi_0^2)^2 +
\frac{1}{2}g^2\phi^2(t)\chi^2 \
\eeq

This has degenerate minima if $\lambda\chi_0^2>g^2\phi(t)^2$. If this is
true at $t=0$, then it is always true; for in the expanding case $\phi$ is
redshifted to zero and in the non-expanding case $\phi$ is not observed to
increase. If we define $\phi_0 \equiv \phi(0)$, then the condition for
degenerate minima becomes: 

\beq \label{cond1}
\lambda\chi_0^2>g^2\phi_0^2 \
\eeq

Another condition follows from the requirement that inflation proceed as
usual, despite the presence of $\chi$. For this we must have the vacuum
energy of $\chi$ to be much smaller than that of $\phi$: 

\beq \label{cond2}
\frac{1}{4}\lambda\chi_0^4 \ll \frac{1}{2}m^2\phi_0^2 \
\eeq
 
When (\ref{cond2}) is combined with (\ref{cond1}), we obtain:

\beq \label{cond12}
g^2\chi_0^2 \ll m^2 \
\eeq 
which ensures the frequency of oscillation of $\phi$ after inflation is
dominated by $m$.
 
We next demand that the maximum reheat temperature be insufficient to lead
to a thermally-corrected potential for $\chi$ with the degeneracy lifted.
Since $T_{RH} < (m\, \phi_0)^\frac{1}{2}$, this condition becomes: 

\beq \label{cond3}
\chi_0 > (\frac{m}{\phi_0})^\frac{1}{2} \phi_0 \
\eeq
Actually the final reheat temperature will be many orders of magnitude
smaller than we have indicated, so (\ref{cond3}) is much stricter than it
needs to be. However it is easy to satisfy and we will retain it in order
to stress we are interested in non-thermal defect production. 
 
Finally there are the usual constraints on $m$ and $g^2$ arising from the
COBE data: $m \leq 10^{-6} M_P$ and $g^2 \leq 10^{-6}$; the latter ensures
radiative corrections do not lead to a too large self-coupling for $\phi$
(though in a supersymmetric model of inflation this condition may not be
necessary \cite{al.pc}).

\vskip 0.5cm \noindent{\bf 2. Analysis}

Our working hypothesis is that parametric resonance will generate large
fluctuations in $\chi$ which will restore the symmetry of the $\chi$-field
in some regions of space. After this, in such regions, $\chi$ may evolve
into the two degenerate minima of its potential, possibly forming a stable
domain wall. In other words, we expect the mechanism of parametric
resonance will be able to channel sufficient energy from the inflaton into
the $\chi$-field in order to drive $\chi$ over the potential barrier. In
this section we outline the conditions for effective parametric resonance,
considering both the expanding and non-expanding cases. 

We begin by assuming the usual flat FRW universe, in co-moving
co-ordinates:

\beqa
ds^2 = dt^2 - a^2(t) (dx^2 + dy^2 +dz^2) \nonumber
\eeqa
 
To a good approximation the coherent oscillations of the inflaton rapidly
give rise to a matter dominated universe so we may write: $a(t) = (1+ m
t)^\frac{2}{3}$, where $t=0$ is the time at the end of inflation and we
have used our freedom to scale the spatial co-ordinates to impose
$a(0)=1$. If we put: $a(t) = (1 + m t)^n$, we can simultaneously consider
the non-expanding case ($n=0$) and the expanding case
($n=\frac{2}{3}$)\footnote{The results of this section are valid only for
these two values of $n$.}.

The equations of motion for the fields $\phi$ and $\chi$ are the standard
ones for scalar fields in a curved space-time with minimal coupling to
gravity: 

\beqa \label{eom1}
\ddot{\phi} + 3H\dot{\phi} - a^{-2}\nabla^2\phi + (m^2 +
g^2\chi^2)\phi = 0 \\
\label{eom2}
\ddot{\chi} + 3H\dot{\chi} - a^{-2}\nabla^2\chi +
(\lambda(\chi^2-\chi_0^2) + g^2\phi^2)\chi = 0
\eeqa
where $H \equiv \frac{\dot{a}}{a}$.
 
These are the equations which are solved in our numerical simulation. Here
we point out that in our simulation the scale factor $a(t)$ is put in by
hand, rather than determined in a self-consistent manner. 
 
The initial conditions for $\phi$ are those of the end of inflation i.e. 
when the slow roll approximation is no longer valid. We find: $\phi(0)
\equiv \phi_0 = \frac{1}{\sqrt{3\pi}}M_P$ and $\dot{\phi}(0)=-m\,\phi_0$.
In addition we chose $m=10^{-6}\phi_0$. For the non-expanding case we will
assume $\dot{\phi}(0)=0$. As for $\chi$, inflation dictates it will be
initially comprised of quantum fluctuations about one of the instantaneous
minima of $V^{eff}_{\chi}$, which are given by $\chi_{\pm}(t) \equiv \pm
\chi_0\,\surd(1-\frac{g^2\phi^2(t)}{\lambda\chi_0^2})$. Without loss of
generality we may choose the positive minimum. 
 
To make headway in our analytical understanding of the model we must make
some simplifying approximations. As already noted, we may take $\phi$ to
be homogeneous to start with. Recalling (\ref{cond12}), this allows us to
solve (\ref{eom1}): $\phi = \phi_0\, a(t)^{-\frac{3}{2}} \cos{mt}$, and to
re-write (\ref{eom2}) as: 

\beq \label{eom3}
\ddot{\chi} + 3H\dot{\chi} - a^{-2}\nabla^2\chi +
\frac{\partial}{\partial\chi}V^{eff}_{\chi} = 0\
\eeq
 
In order to understand the evolution of the $\chi$-field, at least
initially, we will expand $V^{eff}_{\chi}$ to quadratic order about
$\chi_+(t)$:

\beq \label{potapprox}
V^{eff}_{\chi} \longrightarrow U_{\chi} =
\lambda\chi_+^2(t)(\chi-\chi_+(t))^2 +
\frac{1}{4}\lambda(\chi_0^4-\chi_+^4(t)) \
\eeq

How now do we ask if symmetry restoration occurs? We note that
$U_{\chi}(\frac{1}{2}\chi_+(t)) = \frac{1}{4}\lambda\chi_0^4$, which is
the barrier height of $V^{eff}_{\chi}$. Thus the question as to whether
$\chi$ surmounts the potential barrier of $V^{eff}_{\chi}$ may be replaced
with the question: does $\chi$, evolving under the influence of
$U_{\chi}$, ever become less than $\frac{1}{2}\chi_+(t)$? 
 
With the substitution of $V^{eff}_{\chi}$ with $U_{\chi}$ in (\ref{eom3}),
we obtain a linear equation for $\chi$. Upon rescaling: \{$t\rightarrow
\frac{t}{m},\ {\bf x}\rightarrow \frac{{\bf x}}{m},\
\phi\rightarrow\phi_0\phi,\ \chi\rightarrow\phi_0\chi,\ g^2\rightarrow
(\frac{m}{\phi_0})^2\,g^2,\ \lambda\rightarrow (\frac{m}{\phi_0})^2\,
\lambda$\}, taking the Fourier transform and introducing $X_{{\bf k}}
\equiv a^{\frac{3}{2}}\,\chi_{{\bf k}}$, we arrive at a Mathieu equation
with (in general) time-dependent co-efficients, for the
non-zero momentum modes $X_{{\bf k}}$:
 
\beq \label{mathieu}
\ddot{X}_{{\bf k}} + (A_k(t)-2q(t)\cos{2t})\,X_{{\bf k}} = 0
\eeq
where $A_k=\frac{k^2}{a^2}+2\lambda\chi_0^2-\frac{g^2}{a^3}$ and 
$q=\frac{g^2}{2a^3}$.

The existence of exponentially growing, or unstable solutions to the
Mathieu equation is well known\footnote{For a good review see \cite{a64}.}
and, in physics literature, has been termed parametric resonance. However
recent work \cite{k96,kls97,gpr97}, inspired by the pre-heating scenario,
has led to new insight into the nature of these solutions. For
time-independent coefficients, we may identify three different regions of
instability. {\it (i) Narrow band resonance:} $A_k \simeq l^2 \geq 2q$,
$l=1, 2 \ldots$ and $q \ll 1$. The solution $X_{{\bf k}}$ oscillates with
period $2\pi$, and is modulated by an exponential function, $e^{\mu_kt}$,
where $\mu_k$ is called the characteristic exponent or growth index.
Typically $|\mu_k| \sim \frac{q}{2}$. {\it (ii) Broad band resonance:}
$A_k \geq 2q$ and $q \gtrsim 1$. The solution oscillates many times faster
than in the narrow band case and grows in exponential jumps, occurring at
time intervals of $\pi$. The overall growth index may also be much larger:
one finds $|\mu_k| < 0.3$. {\it (iii) Negative instability:} $A_k < 2q$.
The solutions in this region are almost all unstable and $|\mu_k| > 1$ is
possible. Recently Greene et al. \cite{gpr97} have constructed a model
which makes use of this region. 

The situation of the Mathieu equation with time-dependent coefficients is
much more complicated. The difficulty is that, during one oscillation of
the driving force, $X_{{\bf k}}$ does not remain in one instability band
but in fact passes in and out of many such bands. The result is that
parametric resonance becomes essentially haphazard---$X_{{\bf k}}$ may
even decrease in amplitude at times---and hence the term {\it stochastic
resonance} has been applied to this scenario. However exponential type
growth is still possible, and the work of Kofman et al. \cite{kls97} has
done much to clarify this issue. 

Returning to equation (\ref{mathieu}), we see that in our case $A_k$ is
always greater than $q$ because of (\ref{cond1}), hence we will only see
broad band or narrow band resonance. A necessary condition for parametric
resonance\footnote{This is a more generous condition than Eq. (56) in
\cite{kls97}, where $\pi^{-1}$ replaces $1$ on the RHS. We have found, for
our parameter choices, that (\ref{kappa}) includes the (two) main broad
band instability regions.} \cite{kls97} in these regions is: 

\beq \label{kappa}
\kappa^2 \equiv \frac{A_k - 2q}{2\sqrt{q}} \leq 1 \
\eeq

It is useful to introduce the dimensionless parameters: $\alpha=g^2$,
$\beta=\lambda\chi_0^2$ and $\gamma=\chi_0^{-2}$. Then conditions
(\ref{cond1}), (\ref{cond2}) and (\ref{cond3}) may be succinctly written: 
$\alpha < \beta \ll \gamma \lesssim 10^6$. 

For $n=\frac{2}{3}$, $\kappa^2 \rightarrow
\frac{\sqrt{2}\beta}{\sqrt{\alpha}}\,t$ which must always become larger
than $1$. Later we will see that the typical time for symmetry restoration
to begin in the non-expanding case is $t \sim 30$. Thus to satisfy
(\ref{kappa}) in the expanding case we will require $\sqrt{\alpha} \gg
\beta$. But $\alpha < \beta$. So we must have:

\beq \label{timex}
\sqrt{\alpha} \ll 1
\eeq
Or in other words, $g^2_{phys} \ll 10^{-12}$. We conclude that if
(\ref{timex}) is not satisfied, and we believe it is not in any realistic
model, then the system is not long enough in the instability regions to
restore the symmetry. The conclusion is further reinforced if we realise
we also require\footnote{This follows from a simple-minded calculation of
the ratio of energy gained by $X_{{\bf k}}$ to energy lost due to
expansion.} $\mu_k > \frac{3}{2}H$, but that initially $\frac{3}{2}H(0) =
1 > 0.3 \sim \mu_{k,max}$.

The necessary condition for parametric resonance in the case of $n=0$ is
much less stringent. (\ref{kappa}) gives: 

\beq \label{ab} 
\beta \leq \alpha + \sqrt{\frac{\alpha}{2}}
\eeq 

A test of the validity of our approximation in the non-expanding case is
the time taken for symmetry to start being restored. As mentioned
previously, what we need to calculate is the time at which $\chi$ first
becomes less than $\frac{1}{2}\chi_+(t)$. The modes in the resonance band
will dominate the behaviour of $\chi$ so:

\beqa \label{time}
\chi &\approx&
\frac{1}{(2\pi)^{\frac{3}{2}}}\,\frac{m}{\phi_0}\int_{res.\ band}
d^{\,3}k\,\chi_{{\bf k}}\,e^{-i\,{\bf k}.{\bf x}} \nonumber \\
&\sim& \frac{1}{(2\pi)^{\frac{3}{2}}}\,\frac{m}{\phi_0}\, . \,4\pi
\,k^2_{res} \,\frac{\Delta k}{2} \,|\chi_{{\bf k},res}| \nonumber \\
&\approx& \frac{k^2_{res} \,\Delta
k}{(2\pi)^{\frac{1}{2}}}\,\frac{m}{\phi_0}\,\frac{e^{\,\mu_k
t}}{\sqrt{\,\omega_{res}}} \nonumber
\eeqa
where $\Delta k$ is the width of the resonance band and $\omega^2_{res}
\equiv A_{k,res}-2q\cos{2t} = k^2_{res}+2\beta-\alpha-\alpha\cos{2t}$. The
appearance of $\frac{m}{\phi_0}$ and $\sqrt{\,\omega_{res}}$ is due to
$\chi$ starting out as quantum fluctuations \cite{kt97}. 

Typically in our parameter range, we find a region of broad band resonance
for $\beta$ slightly greater than $\alpha$, with $k_{res} \sim \Delta k
\sim 1$ and $\mu_k \sim 0.25$. Then the statement $\chi \leq
\frac{1}{2}\chi_+(t)$ becomes
$\frac{m}{\phi_0}\,\beta^{-\frac{1}{4}}\,e^{\,\mu_kt} \lesssim \chi_0$,
which implies: 

\beq \label{start}
t \gtrsim 4\ln{(10^6\beta^{\frac{1}{4}}\gamma^{-\frac{1}{2}})} \sim 30
\eeq

To summarise: our analysis suggests that in the non-expanding case the
conditions for parametric resonance depend only on $\alpha$ and $\beta$,
whereas the time taken for parametric resonance to restore the symmetry of
the $\chi$-field is determined mainly by $\gamma$. In the expanding case
we do not expect domain walls to be produced. We now turn to the full
simulation of the two-field system. 

\vskip 0.5cm \noindent{\bf 3. Simulation and Results}

The classical equations of motion (\ref{eom1}) and (\ref{eom2}) are a good
approximation to the behaviour of an actual matter field, $\chi$, coupled
to the inflaton, provided the mode amplitudes of $\chi$ are large.
Although initially these are taken to be quantal fluctuations, the modes
we are interested in are in the resonance band and hence they quickly grow
and become classical \cite{kt96}.

We work in terms of the rescaled, non-dimensional quantities of the
previous section. The universe becomes the usual 3-dimensional lattice in
co-moving co-ordinate space, and we apply periodic boundary conditions.
The fields are evolved using a leap-frog scheme, which is an explicit
algorithm second-order accurate in time. The spatial derivatives are
evaluated to fourth-order accuracy.

The results presented here were obtained on $64^3$ lattices, though we
also checked representative cases on $128^3$ lattices. We varied the
lattice spacing, the only proviso being that we remained sensitive to the
momentum modes in the expected resonance bands, and obtained essentially
identical data. The total energy of the system was conserved to better
than $0.1\%$. Two other tests convinced us of the accuracy of our
simulation: we were able to repeat the results of \cite{pr97}, and, in an
$O(2)$ $\chi$-field simulation, we found a faithfully modelled, collapsing
string loop. 

A brief note about the determination of the initial conditions for $\chi$
is warranted. As mentioned before, we treat $\chi$ initially as quantum
fluctuations about one of the instantaneous minima of its potential. One
may think of the quantum fluctuations in momentum space as an infinite
collection of harmonic oscillators (with time dependent frequencies in
this case) \cite{b84}. Our approach essentially was to pick the occupation
numbers of the modes on the lattice randomly from a Gaussian distribution
with $k$-dependent width \cite{kt97}. We also included a random phase and
then inverse Fourier transformed to obtain $\chi$ in configuration space.
We checked that our results were not sensitively dependent on the
particular random numbers chosen. A final important point is that the
lattice spacing gives a natural cutoff to the high momentum modes, and one
must choose this so that $E_{fluct.} \ll E_{inflaton}$. 

We examined the region in parameter space of $\alpha, \beta \sim 100$ and
$\gamma \sim 10^4$; in other words $g_{phy}^2 \sim 10^{-10}, \lambda_{phy}
\sim 10^{-6}$ and $\chi_{0,phy} \sim 0.01\phi_0 \approx 3 \times 10^{16}$
GeV. The choice of these parameters was partially determined by the
requirements that we be sensitive to the instability bands and the box be
large enough to include any defects produced \cite{ps98b}. The numerical
parameters were: lattice spacing $\Delta x = 0.1-1$ and time step $\Delta
t = 0.02$. 

The first result is that we did not find any evidence of symmetry
restoration in an expanding universe. In all cases we considered, $\chi
\rightarrow \chi_0$ and $\phi$ merely redshifted to 0. This is as
predicted by (\ref{timex}) and Kofman et al. \cite{kls96}.

\begin{figure}
\mbox{\epsfig{figure=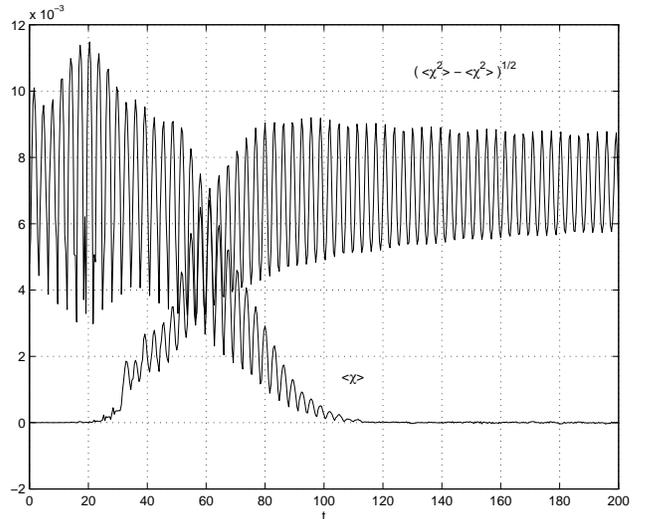,height=7cm}}
\caption{The mean and variance of $\chi$ in time ($\alpha=100$,
$\beta=130$ and $\gamma=10^4$).}
\end{figure}

The non-expanding scenario is more interesting however. Figure 1 shows the
behaviour of $\langle\chi\rangle$ and
$\surd(\langle\chi^2\rangle-\langle\chi\rangle^2)$ for $\alpha=100$,
$\beta=130$ and $\gamma=10^4$. This is typical of the cases in which
symmetry is restored: the mean value of $\chi \rightarrow 0$ and the
fluctuations grow to be of the order of $\chi_0$. Also typical is the fact
that the back-reaction of $\chi$ on $\phi$ is very slight.
$\langle\phi\rangle$ continues to oscillate sinusoidally and the
fluctuations of $\phi$ are only of order $3 \times 10^{-3} \ll 1$ at
$t=200$. 

\begin{figure}
\mbox{\epsfig{figure=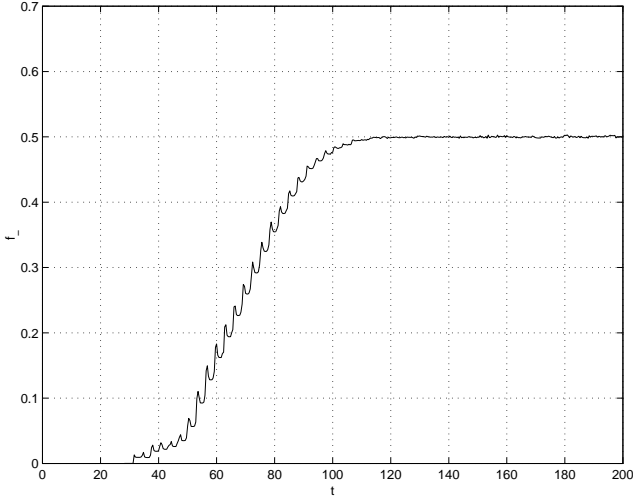,height=7cm}}
\caption{The fraction of the volume with $\chi<0$ ($\alpha=100$,
$\beta=130$ and $\gamma=10^4$).}
\end{figure}

Plotted in figure 2 is the fraction, $f_-$, of the volume of the box with
$\chi <0$. The jumps in $f_-$ occur at intervals of $\pi$ which is as
expected; $\pi$ is the period of the driving term in the Mathieu equation
i.e. $\phi^2$. Note too the onset of symmetry restoration is at $t \approx
30$, confirming our earlier order of magnitude estimate. However,
considering all our runs, we did not find agreement with the specific form
of (\ref{start}). (Usually $t = 15-80$.) On the other hand varying
$\gamma$, for fixed $\alpha$ and $\beta$, did not alter whether symmetry
was or was not restored, as was our prediction. 

In the non-expanding case, symmetry restoration (when it occurs) gives
rise to non-thermal production of defects. However these defects are not
stable. This can be seen in our simulations: the regions of $\chi < 0$ are
essentially randomly scattered throughout the box (even when they account
for half the volume of the box) and percolate. In addition they are
uncorrelated in time. 

\begin{figure}
\mbox{\epsfig{figure=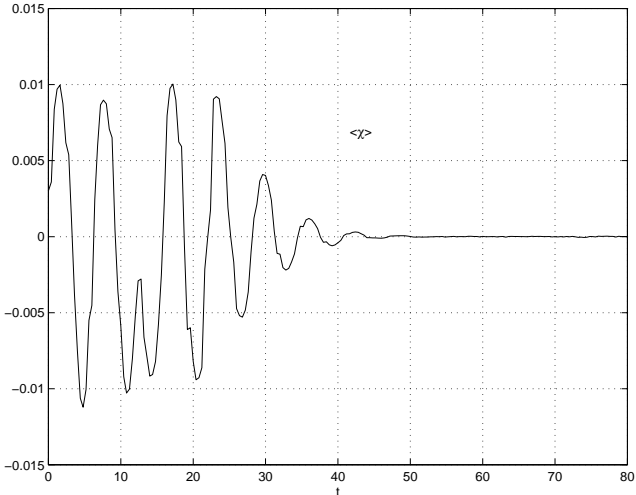,height=7cm}}
\caption{The mean of $\chi$ in time ($\alpha=100$,
$\beta=110$ and $\gamma=10^4$).}
\end{figure}

There is one problem with the particular case we have been considering
above: it does not satisfy (\ref{ab}). Thus the analysis of section 2
cannot be entirely correct. Figures 3 and 4 detail what was an unexpected
result but which led to a more complete understanding of our model. Here
$\alpha = 100$, $\beta = 110$ and $ \gamma = 10^4$. Clearly what is
happening is an initial series of global sign changes, prior to eventual
local symmetry restoration. The change in sign is not important because
the symmetry of $\chi$ remains broken, but it does indicate we were naive
to imagine $\chi$ would fluctuate about $\chi_+(t)$ for early times. 

\begin{figure}
\mbox{\epsfig{figure=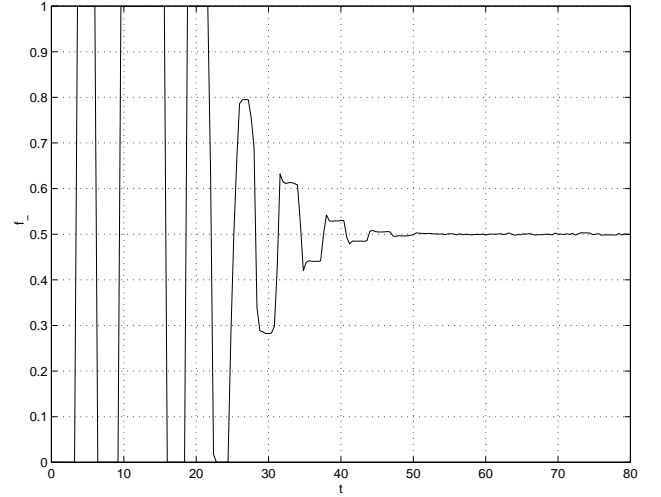,height=7cm}}
\caption{The fraction of the volume with $\chi<0$ ($\alpha=100$, 
$\beta=110$ and $\gamma=10^4$).}
\end{figure}

It is more appropriate to put $\chi({\bf x},t) = \chi_h(t) +
\delta\chi({\bf x},t)$, where $\chi_h$ satisfies the homogeneous version
of (\ref{eom3}). Then the linearised equation for $X_{{\bf k}} \equiv
a^{\frac{3}{2}} \delta\chi_{{\bf k}}$, which replaces (\ref{mathieu}), is: 
 
\beq \label{eomd}
\ddot{X}_{{\bf k}} + (\frac{k^2}{a^2} +
\lambda(\,3\chi_h^2(t)-\chi_0^2)+g^2\phi^2(t))\,X_{{\bf k}} =
0
\eeq
 
The behaviour of $\chi_h$ is quite delicate. Numerical solution of its ODE
shows that for early times it is true $\chi_h \approx \chi_+$, {\it
except} when $\beta \approx \alpha$ where we find $\chi_h \approx
\pm\chi_+$. This is the sign change seen in figures 3 and 4. At later
times $\chi_h^2$ typically oscillates about $\chi_+^2$, with a frequency
several times greater than that of the inflaton. For $n=\frac{2}{3}$,
$\chi_h^2 \rightarrow \chi_+^2 \rightarrow \chi_0^2$\, so in this case we
believe our earlier analysis is essentially unchanged. However for $n=0$,
$\chi_h^2$ tends to increase in amplitude. This ruins a straight-forward
analysis in terms of a Mathieu equation for $X_{{\bf k}}$ (which would
come about if $\chi_h^2 = \chi_+^2$). Numerical analysis of (\ref{eomd})
for $n=0$ though does reveal exponential-type solutions. In fact
parametric resonance appears to be much more likely: the instability bands
are larger than those of the broad band regime of the Mathieu equation. We
are currently working on an analytical understanding of these
results\cite{ps98b}. 

\begin{figure}
\mbox{\epsfig{figure=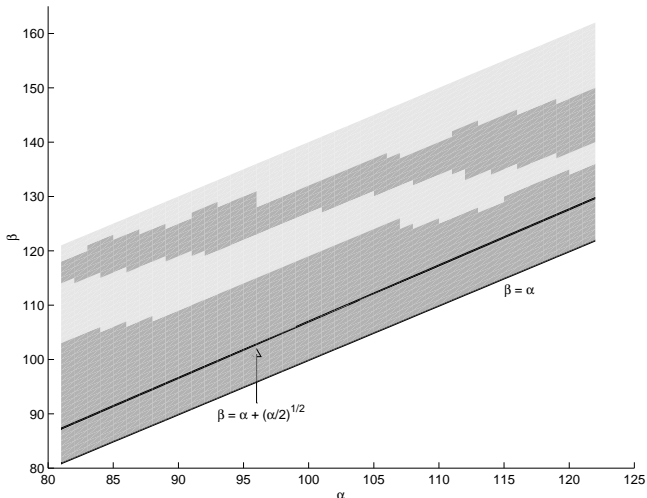,height=7cm}}
\caption{The region of symmetry restoration (darker shading) in the
$\alpha-\beta$ plane.}
\end{figure}

To conclude this section figure 5 shows the region of the $\alpha-\beta$
plane we investigated where symmetry restoration occurs, for the
non-expanding case. We stress that this plot is an extrapolation from 45
simulations we performed, and as such to be taken only as a rough guide.
However it indicates two features which we are certain about. Firstly the
existence of band structure. This is reminiscent of the Mathieu equation
instability bands, even down to their relative sizes. Secondly the fact
that as $\beta$ increases for fixed $\alpha$, symmetry restoration is no
longer possible. From this graph we conservatively deduce a condition for
symmetry restoration: 

\beq \label{afew}
\beta < 2\,\alpha
\eeq

Also shown in figure 5 is our prediction from section 2. The discrepancy
is the most striking evidence that our earlier analysis was inadequate.

\vskip 0.5cm \noindent{\bf 4. Conclusions}

We have numerically investigated the decay of the inflaton into a scalar
matter field $\chi$ with a symmetry breaking potential, when the symmetry
is broken during inflation. In the non-expanding case we found that
symmetry restoration is possible via parametric resonance. The analysis is
complicated but interestingly the regions of instability are much larger
than those of the Mathieu equation. This situation of more efficient
parametric resonance has also been seen in the work of Zanchin et al.
\cite{zmcb97}. We think it would be useful to more fully understand
equations like (\ref{eomd}), perhaps utilising the methods of
\cite{kls97}. Finally we found that non-thermal defect production occurs,
but that the defects are not stable.

In the expanding case we found that domain wall production did not take
place. This leads us to conclude that the inflaton may safely couple to
$\chi$ with the potential as in (\ref{pot}). Actually the precise form for
the potential may be unimportant. Recall that the symmetry of $\chi$ is
broken before the end of inflation. In fact, if there are to be no
fluctuations due to the initial symmetry breaking in our Hubble volume,
the symmetry must be broken early enough to allow for the usual number of
$e$-foldings by the time inflation is over. In our model this means $\beta
\gtrsim 100\,\alpha$\cite{ps98b} which is to be compared with
(\ref{afew}). This amounts (literally) to a very formidable barrier to
symmetry restoration. This suggests it is unlikely that there will be
defect production in any realistic model in which $\chi$ has the symmetry
breaking potential and the symmetry is broken during inflation.

\vskip 0.5cm \noindent{\bf Acknowledgements}

It is a pleasure to be able to thank Tomislav Prokopec, Richard Easther,
Guy Moore and Martin G\"{o}tz who were helpful at crucial stages. In
particular we would like to thank Anne-Christine Davis who initially
jump-started this investigation, Robert Brandenberger who was a continual
source of help, and Andrei Linde who more than once gave one of us (MP) 
pertinent advice. Computational work in support of this research was
performed at the Theoretical Physics Computing Facility at Brown
University. This work was supported by the DOE and the NASA grant NAG
5-7092 at Fermilab.


\begin{references}

\bibitem[*]{mail1}Email: parry@het.brown.edu

\bibitem[\dagger]{mail2}Email: ats@traviata.fnal.gov 

\bibitem{tb90} J. Traschen and R. Brandenberger, Phys. Rev. D{\bf 42}, 2491
(1990)

\bibitem{kls94} L. Kofman, A. D. Linde and A. A. Starobinskii, Phys. Rev. 
Lett. {\bf 73}, 3195 (1994)

\bibitem{stb95} Y. Shtanov, J. Traschen and R. Brandenberger, Phys. Rev. 
D{\bf 51}, 5438 (1995) 

\bibitem{krt98} E. W. Kolb, A. Riotto and I. I. Tkachev, hep-ph/9801306

\bibitem{kls96} L. Kofman, A. D. Linde and A. A. Starobinskii, Phys. Rev. 
Lett. {\bf 76}, 1011 (1996) 

\bibitem{t96} I. I. Tkachev, Phys. Lett. B{\bf 376}, 35 (1996)

(1997)

\bibitem{kklt98} S. Khlebnikov, L. Kofman, A. Linde and I. Tkachev,
hep-ph/9804425

\bibitem{kk98} S. Kasuya and M. Kawasaki, hep-ph/9804429

\bibitem{kl87} L. Kofman and A. D. Linde, Nucl. Phys. B{\bf 282}, 555
(1987)

\bibitem{al.pc} A. D. Linde, {\it private communication}

\bibitem{a64} F. M. Arscott, ``Periodic Differential Equations'', (The
MacMillan Company, New York 1964)

\bibitem{k96} L. Kofman, astro-ph/9605155

\bibitem{kls97} L. Kofman, A. D. Linde and A. A. Starobinskii, Phys. Rev.
D{\bf 56}, 3258 (1997)

\bibitem{gpr97} B. Greene, T. Prokopec and T. G. Roos, Phys. Rev. D{\bf
56}, 6484 (1997)

\bibitem{kt97} S. Yu. Khlebnikov and I. I. Tkachev, Phys. Lett. B{\bf
390}, 80 (1997) 

\bibitem{kt96} S. Yu. Khlebnikov and I. I. Tkachev, Phys. Rev. Lett. {\bf
77}, 219 (1996)

\bibitem{pr97} T. Prokopec and T. G. Roos, Phys. Rev. D{\bf 55}, 3768
(1997)

\bibitem{b84} R. Brandenberger, Nucl. Phys. B{\bf 245}, 328 (1984)

\bibitem{zmcb97} V. Zanchin, A. Maia, Jr., W. Craig and R. Brandenberger,
Phys. Rev. D{\bf 57}, 4651 (1998)

\bibitem{ps98b} M. F. Parry and A. T. Sornborger, {\it in preparation}

\end{references}
\end{document}